\documentclass[12pt,preprint]{aastex}
\newcommand{\RNum}[1]{\uppercase\expandafter{\romannumeral #1\relax}}

\shorttitle{emerging active region, fan-shaped activities}
\shortauthors{Zhong et al.}

\begin{document}

\title{The dynamics of AR 12700 in its early emerging phase \RNum{2}: fan-shaped activities relevant to arch filament systems}

\author{Sihui Zhong\altaffilmark{1,2}, Yijun Hou\altaffilmark{1,2},
Leping Li\altaffilmark{1,2}, Jun Zhang\altaffilmark{1,2}, and Yongyuan Xiang\altaffilmark{3}}

\altaffiltext{1}{CAS Key Laboratory of Solar Activity, National Astronomical Observatories,
Chinese Academy of Sciences, Beijing 100101, China; zsh@nao.cas.cn; zjun@nao.cas.cn}

\altaffiltext{2}{University of Chinese Academy of Sciences, Beijing 100049, China}

\altaffiltext{3}{Yunnan Observatories, Chinese Academy of Sciences, Kunming 650011, China}
\begin{abstract}
The emergence of active regions (ARs) closely relates to the solar dynamo and the dynamical atmospheric phenomena.
With high-resolution and long-lasting observations from the New Vacuum Solar Telescope, we report a new dynamic activity phenomenon named
``fan-shaped activity (FSA)" in the emerging phase of NOAA AR 12700.
The FSAs are clearly observed at H$\alpha$ wavelength and are closely related to the dynamics of the adjacent arch filament system (AFS),
including threads deformation and materials downward motions. On 2018 February 26, the two most representative FSAs appeared around 05:21 UT and 06:03 UT, respectively, and they firstly ascended and then decayed in around 10 minutes.
At the ascending phase,
accompanied by the uplifting of an adjacent AFS, each FSA launches up at one end of the AFS and extends for up to $\sim$11 Mm.
At the decaying phase, the FSA gradually vanishes, and materials downflows towards the other end of the AFS are detected.
After checking the evolution of the magnetic fields of AR 12700, we find that each FSA is located between the end of an AFS and an adjacent magnetic patch
with the same polarity and launches at the onset of the collision and compression between these two magnetic patches.
We propose that the collision lifts up the AFS, and then the initially compact AFS laterally expands,
resulting in the formation of FSA. A cartoon model is proposed to depict the activities.
\end{abstract}


\section{Introduction}
The emergence of active regions (ARs) is a multi-stage process along with various complex phenomena \citep{2015LRSP...12....1V}.
Magnetic flux tubes emerging through the photosphere will form a series of bipoles, and then the conjugated polarities
of each bipole move apart. Eventually patches with the same polarity may merge to form pore-like features. When rising to
the chromosphere, the emerging flux tubes are traced out by arch filament system (AFS; \citealt{1967SoPh....2..451B}).
AFS is dark arch-shaped feature connecting two opposite magnetic polarities of newly emerging bipoles, with an upward motion
at their tops and plasma downward motion along their legs. \cite{2004A&A...425..309S} observed the AFS dynamic evolution
during the emergence of AR 10050, which showed that the values of both upflow and downflow velocities of the AFS decreased during the evolution of the AR.

Recently, the high temporal and spatial resolution observations have lead to significant progress on the research about AR emergence.
Data from \emph{Hinode}/Solar Optical Telescope (SOT; \citealt{2008SoPh..249..167T}) have revealed the nature of flux emergence
\citep{2011PASJ...63.1047O} and evidenced granular-scale elementary flux emergence episodes during the emergence of AR 11024 \citep{2012SoPh..278...99V}. Observations from \emph{Hinode}, \emph{Interface Region Imaging Spectrometer}
\citep{2014SoPh..289.2733D}, and \emph{Solar Dynamics Observatory }(\emph{SDO}; \citealt{2012SoPh..275....3P})
detected various local heating events in a new emerging-flux region of AR 12401 \citep{2017ApJ...836...63T}.
Several ground-based solar telescopes have also made great contributions. Employing observations of the Vacuum Tower Telescope (VTT; \citealt{1998NewAR..42..493V}), \cite{2010A&A...520A..77X} performed the first investigation on the vector magnetic field and Doppler velocity in the lower solar atmospheric layers of a young emerging-flux region. \cite{2017A&A...600A..38G} analyzed VTT observations of an AFS and presented the decay and convergence of two micro-pores with
diameters of less than one arc-second in a small emerging-flux region.
Goode Solar Telescope
\citep{2012ASPC..463..357G} observed the emergence of small-scale flux ropes in the photosphere \citep{2014ApJ...794..140V}
and the chromosphere \citep{2017ApJ...845...18Y}.
Moreover, several studies of AFS based on spectropolarimetric observations by GREGOR \citep{2012AN....333..796S}
reported supersonic downflows along the legs of the arch filaments \citep{2016AN....337.1050B,2018A&A...617A..55G}.
In our previous work
\citep{2019ApJ...876...51Z}, combining observations from the New Vacuum Solar Telescope (NVST; \citealt{2014RAA....14..705L})
and the nonlinear force-free field (NLFFF) extrapolation results, the detailed process of interchange reconnections in the center of AR 12700 during its emerging phase were presented.

Great efforts at numerical simulations have also been made to model the flux emergence (see review of \citealt{2014LRSP...11....3C} and the references therein),
and recent three-dimensional (3D) magnetohydrodynamics simulations have successfully produced an AR based on different emergence conditions
(\citealt{2010ApJ...720..233C,2012A&A...537A..62A, 2014ApJ...785...90R, 2017ApJ...846..149C}; and reference therein).
Additionally, combining direct analysis of the numerical data, 3D visualization, and spectropolarimetric synthesis,
\cite{2018ApJ...859L..26M} identified two types of small-scale emerging magnetic structures in the quiet Sun.
The emerging granule-covering flux sheets in their models may match the subarcsecond resolution observations of \cite{2017ApJS..229....3C}.

When magnetic flux emerges from beneath the photosphere, it may reconnect with the preexisting fields, which would heat the local plasma,
give rise to jets and emit waves that propagate into the corona \citep{2008ApJ...679L..57I}.
Surges, the most common
dynamic activities in solar atmosphere, are ejections of cool and dense plasma in the chromosphere.
Generally, they eject upwards along
straight or curved paths at velocities of 10 to 200 km s$^{-1}$, reaching heights of 5 to 100 Mm, and lasting for 10 to 30 minutes
\citep{1982SoPh...81....9S, 1995SoPh..156..245S, 1996ApJ...464.1016C}. Observationally, surges are often found at emerging flux region
in its earliest stage \citep{1993ASPC...46..507K}, or on light bridge in the well-developed or decaying sunspots
\citep{ 2014A&A...567A..96L, 2016A&A...590A..57R, 2017ApJ...848L...9H,2018ApJ...854...92T}.
It is widely believed that
the surge is caused by magnetic reconnection related to flux emergence or flux cancellation, which is supported by observational evidences
\citep{1999ApJ...513L..75C, 2003PASJ...55..313Y, 2004ApJ...610.1136L, 2007MNRAS.376.1291B, 2016ApJ...826..217L}
and numerical simulations \citep{1995Natur.375...42Y,1996PASJ...48..353Y, 2011ApJ...726L..16J, 2012ApJ...751..152J, 2017ApJ...850..153N}.

In this paper, we report a new dynamic activity phenomenon, fan-shaped activity (FSA), in the emerging phase of AR 12700 on 2018 February 26.
This kind of activities is related to AFS and owns a similar appearance to surge but without brighteness in extreme ultraviolet (EUV) channel.
We analyse the characteristics and driver mechanism of FSA based on the H$\alpha$ observations acquired at the NVST
and the simultaneous observations from Helioseismic and Magnetic Imager (HMI; \citealt{2012SoPh..275..207S}) as well as Atmospheric Imaging Assembly (AIA; \citealt{2012SoPh..275...17L}) on board the \emph{SDO}.
At H$\alpha$ wavelength, several FSAs are detected, which are closely associated with the dynamics of the adjacent AFS,
including threads deformation and materials downflows. The two most representative FSAs appeared around 05:21 UT and 06:03 UT,
respectively. They exhibits a rapid evolution, ascending and decaying within a period of about 10 minutes. At the ascending phase, accompanied
by the uplifting of the adjacent AFS, each FSA rises at one end of the AFS and extends for up to $\sim$11 Mm.
After that,the FSA fades away, with materials falling down towards the other end of the AFS.

Our paper is organized as follow. In Section 2, we describe the observations and data analysis taken in this
study. Section 3 investigates the FSA in detail. Finally, we summarize the major findings, discuss the results,
and suggest a possible mechanism  for the formation of the FSA in Section 4.

\section{Observations and Data Analysis}
The emerging AR 12700 with $\beta-$configuration at solar disk location N04W01 was observed by the NVST on 2018 February 26.
The instruments currently installed on NVST include the adaptive optics (AO) system, multi-channel imaging system, and two vertical grating spectrometers \citep{2014RAA....14..705L}.
The imaging system of NVST consists of one channel for the chromosphere (H$\alpha$) and two channels for the photosphere (TiO and G-band).
The NVST H$\alpha$ 6563 {\AA} observations adapted in this work were taken from 02:01:00 UT to 06:56:00 UT and covered a field of view (FOV) of 152{\arcsec} $\times$ 151{\arcsec} with a pixel size of 0.{\arcsec}136 and a cadence of 8 s.
The NVST H$\alpha$ images can clearly display the detailed AR emergence process in the chromosphere, including fibrils emergence, reconnections between different groups of fibrils, and FSAs. In this work, the two representative FSAs appearing at 05:21 UT and 06:03 UT are studied in detail.
In order to figure out the photospheric magnetic field evolution and coronal response during the activities, we have also analyzed the data taken by the HMI and AIA on board \emph{SDO}.
The HMI data adopted here were obtained from 2018 February 25 to 26. The time resolution of the line-of-sight (LOS) magnetograms taken by HMI is 45 s and the spatial resolution is 1.{\arcsec}0.
The AIA provides simultaneous full-disk images of the multi-layered solar atmosphere with 10 passbands, 7 of which are in EUV channel and observed with a cadence of 12 s and a pixel size of 0.{\arcsec}6.
All images observed by the \emph{SDO} are differentially rotated to a reference time (04:00:00 UT on February 26).
Moreover, data from all telescopes and instruments are carefully co-aligned by using the AIA images as a reference.
According to some features which can be simultaneously detected in SDO/AIA 171 {\AA}, and NVST Ha channels, such as dark fibrils and bright points, the original H$\alpha$ images are rotated and shifted to be co-aligned with the AIA images. Since AIA images and HMI magnetograms have already been co-aligned, HMI magnetograms and the processed H$\alpha$ images are then co-aligned as well.

To derive the horizontal photospheric velocities during the two FSAs, we apply the differential affine velocity estimator
(DAVE; \citealt{2006ApJ...646.1358S}).
The DAVE method combines the magnetic induction equation and an affine velocity profile to a windowed aperture of
the magnetogram sequence to determine the optical flow of magnetic footpoints.
We set the window size as 12 pixels in DAVE,
and obtain the velocity field from
the difference between two sets of HMI LOS magnetograms with five-minute interval.

In addition, we make time-slice plot to obtain the velocity of the downflow. Firstly, we approximate the trajectory
of downward material flow by an arc-sector domain. During the period of the downward motion, we collect the imaging data within
the arc-sector domain and store them as a two-dimensional array every 8 seconds. Then we merge these arrays into one larger
two-dimensional array in chronological order. Finally, we display the final array in a figure (time-slice plot) with its X-axis
referring to time and Y-axis to distance. In this manner, the slope of the dashed line connecting each endpoint of
the downflow in the time-slice plot indicates the projective downward velocity.

In order to investigate the magnetic configuration of the two FSAs, we use the ``weighted optimization"
method \citep{2004SoPh..219...87W, 2012SoPh..281...37W} to perform NLFFF extrapolations at 05:24 UT and 06:12 UT on February 26, respectively.
We utilize the HMI photospheric vector magnetic fields as the boundary condition. Here, the vector magnetograms are preprocessed
by a procedure developed by \cite{2006SoPh..233..215W} to satisfy the force-free condition. Both NLFFF extrapolations are performed in
a box of 288 $\times$ 168 $\times$ 256 uniformly spaced grid points (104 $\times$ 61 $\times$ 93 Mm$^{3}$).

\section{Results}
With the high-resolution and long-lasting H$\alpha$ observation of NVST from 02:01 UT to 06:56 UT on February 26, we detect a new interesting phenomenon within the emerging AR 12700,
namely FSA.
Several FSAs are observed, which have the similar appearance to the H$\alpha$ surges reported in \cite{2009ApJ...696L..66S} and \cite{2016A&A...590A..57R}. And two FSAs are investigated in detail in the present work (see Figure 1).
Each FSA is located at the footpoint of an AFS, which is detected clearly at EUV and H$\alpha$ wavelengths.

The first fan-shaped activity (FSA1) was well observed from 05:21 UT to 05:\textbf{35} UT on February 26 (see Figure 2 and animation 1),
which was located at the north end of an AFS (AFS1). The northern footpoint of AFS1 was anchored in a positive magnetic patch
as shown in panel (a). FSA1 ascended from 05:21 UT to 05:28 UT, peaked at 05:29 UT, and descended from 05:30 UT to 05:35 UT.
At the ascending stage of FSA1, small-scale threads spouted out at the north end of AFS1 and swept from west to east,
which were projected onto a surface to exhibit a fan-shaped feature.
At the peak of the activity (Figure 2(d)), the fan-shaped feature consisted of three threads (which are outlined by white dotted curves)
with a maximum projected length of about 5 Mm. At the descending stage of FSA1, shown as panel (e), distinct downflow
was detected from 05:31 UT to 05:34 UT at the other footpoint of AFS1. The averaged velocity of the flow towards the south end was $\sim$\textbf{33} km s$^{-1}$.

To investigate the origin of this FSA, we check the photospheric magnetic field evolution of AFS1 (see figure 3 and animation 2).
The bipole connected by the AFS1 initially emerged at about 02:00 UT. Then the positive and negative patches of this bipole drifted northeastward and southwestward, respectively.
In the following three hours, these two target patches constantly merged with the surrounding magnetic elements with the same polarities, finally forming the two polarities (P1/N1) connected by AFS1.
From 05:00 UT, P1 drifted towards the north patch (P2) and then the collision between P1 and P2 occurred at where the FSA subsequently launched. In addition, we calculated the photospheric velocity fields by analyzing the HMI LOS magnetograms with DAVE method.
As shown in panels (b1)-(b2), at 04:12 UT, P1 was approaching P2 with the velocity of $\sim$0.45 km s $^{-1}$. And at 05:13 UT, the approaching velocity rose to $\sim$0.5 km s$^{-1}$, after which compression and collision between P1 and P2 occurred.
Combining the HMI magnetograms and H$\alpha$ observations, it is clearly shown that FSA1 launched soon after the collision started.

The second fan-shaped activity (FSA2) was well observed from 06:03 UT to 06:28 UT (see Figure 4 and animation 3), which launched near the southwestern footpoint of another AFS (AFS2).
FSA2 ascended from 06:03 UT to 06:13 UT, peaked at about 06:14 UT, and descended from 06:15 UT to 06:20 UT.
At 06:03 UT, accompanied by the uplifting of the AFS2, thread-like features began to fan out, forming a fan-shaped feature.
At the ascending stage, the number of the threads that make up the fan-shaped feature grew and their length extended, with the maximum projected length up to $\sim$11 Mm.
After that, at the northeastern end of AFS2, materials falling at a velocity of  $\sim$32 km s$^{-1}$ were clearly detected from 06:15 UT to 06:16 UT (see panel (d)).

Figure 5 displays the photospheric magnetic field evolutions of AFS2 (also see animation 4). The bipole connected by the AFS2
emerged around 18:00 UT on February 25 and the emergence region is outlined by the blue parentheses in Figure 4(a1).
Then the bipole connected by the third AFS (AFS3) emerged at 00:07 UT on February 26. They all separated from each other
along northeast-southwest direction. In the following couples of hours, these bipoles constantly merged with the surrounding magnetic elements,
finally forming the polarities (P3/N3 and P4/N4 in panels (a2)-(a3)) connected by AFS2 and AFS3.
Note that P3 combined with P4 as one big magnetic patch soon after their emergence.
As shown by the HMI magnetograms
and velocity fields in panels (b1)-(b2) at 05:24 UT, N3, N4, and N moved westwards at velocities of 0.23 km s$^{-1}$,
0.22 km s$^{-1}$, and 0.21 km s$^{-1}$, respectively. And at 06:13 UT, the velocities rose to 0.38 km s$^{-1}$, 0.59 km s$^{-1}$,
and 0.29 km s$^{-1}$, respectively. This indicates that N3 was pushed by N4 to collided with N. Combining the HMI magnetograms
and H$\alpha$ observations, FSA2 was located between N3 and N, and launched at the onset of the collision between these two magnetic patches.

Based on the photospheric vector magnetic fields at 05:24 UT and 06:12 UT, we extrapolate the 3D magnetic fields of AR 12700
by using NLFFF modeling. Figures 6(a1) and 6(a2) show the extrapolation results at 05:24 UT
from the top view and side view, respectively.
It is shown that a set of magnetic field lines connect P1 and N1, corresponding to AFS1, see the green curves in Figures 6((a1)-(a2)). To the north of P1,
a bundle of nearly open field lines are rooted at another positive magnetic patch (P2). Figures 6(b1) and 6(b2) display that, at 06:12 UT,
another set of magnetic field lines connect the dipole (N3/P3) located in the inner AR, representing AFS2, see the green curves in Figures 6((b1)-(b2)). Adjacent to N3,
a set of magnetic field lines with higher altitude are located above a negative magnetic patch (N), see the pink curves in Figures 6((b1)-(b2)). To some extent, the extrapolation results
are consistent with the observations shown in Figures 2--5.

\section{Summary and Discussion}
In this paper, we study a new dynamic activity phenomenon, i.e., FSA, which appeared in emerging AR 12700 on February 26,
by using observations from the NVST and \emph{SDO}.
The two most representative FSAs appeared around 05:21 UT and 06:03 UT, repectively. Each FSA manifested as fan-shaped feature locating at the footpoint of an adjacent AFS,
extending for up to 11 Mm and lasting for 10-35 minutes. At the decaying phase of the activities, the fan-shaped features gradually vanished and distinct materials downward motion
was detected on the other end of the relevant AFS.
Moreover, photospheric magnetic fields evolution and the horizontal photospheric velocity fields showed that these two FSAs might be associated with the compression and collision between one end of the relevant AFS and an adjacent magnetic patch with the same polarity.

As shown in Section 3, the observed FSAs are similar to surges but actually not surges. Firstly, the plasma of the fan-shaped feature didn't ejected from the footpoints as the typical surges do but showed up in the midway.
Secondly, observations in the present work haven't detected any eruption or brightenings accompanying the FSAs while typical surges usually coexist with multiwavelength brightenings \citep{2004ApJ...610.1136L,2008ApJ...683L..83N}.
Thirdly, the sequence of magnetograms and velocity fields showed that the observed fan-shaped features were related to the compression and collision of different patches with the same polarity.
Previous reported H$\alpha$ surges or chromospheric jets usually are located at the polarity inversion line \citep{2012ApJ...752...70U} or mixed polarity region \citep{2007Sci...318.1591S,2015ApJ...811..137T}.

There may be several possible driver mechanisms of the FSA. In this study we propose two possible interpretations for the FSA formation, one of which is that these two FSAs may result from the compression and collision between two magnetic patches with the same polarity.
We check the photospheric magnetic field evolution during the FSA1 and find that the northern end of AFS collided with the northern patches with the same polarity at around 05:16 UT, likely forming an magnetic interface.
The magnetic interface was probably where the FSA1 lain. Likewise, the FSA2 was also associated with the collision between two negative polarities.
In addition, at the decaying phase of the FSAs, in the upper atmosphere shown by H$\alpha$ observations, we find that materials drained down along the legs of AFS with the projected velocity up to 33 km s$^{-1}$, which is comparable to the observations in \cite{2018A&A...617A..55G}. Previous H$\alpha$ observations focusing on AFS detected downflows at their endpoints with velocities in the range of 10-50 km s$^{-1}$ \citep{1969SoPh....8...29B,2004A&A...425..309S,2005A&A...442..661Z}.
The downflow indicates the effect of gravity, supporting the idea that the AFS, which is carrying the materials, has been lifted up.
Combining the observations and the NLFFF extrapolation results, we interpret the FSAs as a result of the deformation and uplifting of one leg of AFS,
which are triggered by the collision and compression between the end of the AFS and an adjacent magnetic patch with the same polarity.

To better illustrate this collision model, we sketch several cartoons for the case of FSA1 in Figure 7. At the beginning,
the AFS is stable, connecting the opposite polarities (P1 and N1) of a bipole. Then, the positive patch (P1) of AFS shifts
northeast and collides with the northernmost positive patch (P2). Therefore, some magnetic field lines of AFS are deformed
and lifted up by the magnetic compression. At the north end of AFS, the initially compact AFS expands laterally, shaping a fan-shaped feature.
Another factor contributing to the FSA formation may be that the plasma are blocked by magnetic interface, then piled up,
and the density is increased hence visible at H$\alpha$ wavelength. In addition, materials in the fields of AFS fall down to the south end (N1) because of the action of gravity.

Photopsheric magnetic properties including flux emergence, flux cancellation, magnetic reconnection and shearing are
closely associated with different solar activities \citep{2001ApJ...548L..99Z,2007ApJ...662L..35Z,2012SoPh..278...33V,2015ApJ...809...34C,2017A&A...603A..36K,2018A&A...619A.100H}. Recent observations of \cite{2019ApJ...871...67C}
revealed that collisional shearing, a process during which the collision between nonconjugated polarities of multiple bipoles
results in shearing and flux cancellation, leads to major solar activities. In the present work, the collision between the same
polarities and the effects of this collision are investigated. This type of collision may drive the dynamic evolution of AFS,
and then result in the formation of FSA, which were observed for the first time.

In addition, FSA may also be the result from the reconnection occurring between the AFS and its overlying fields.
Previous studies have revealed that emerging flux tube can reconnect with the preexisting magnetic fields, giving rise to
jets or flares \citep{2010ApJ...724.1083G,2014ApJ...794..140V,2017ApJ...845...18Y}. \cite{2018ApJ...855...77S}
reported that five sequential flares are caused by 3D magnetic reconnection at the quasi-separatrix layers associated with
the AFS during the emerging phase of AR 12396. In our case, there are no brightenings detected in EUV and H$\alpha$ channels.
Thus the reconnection here might be weak with a small reconnection rate. \cite{2008A&A...488.1117Z} argued that
they observed weak reconnection between a new emerging magnetic flux system and the preexisting fields because the
relative orientation of these two flux systems is $\sim$12$^{\circ}$. Moreover, \cite{2007ApJ...666..516G} presented similar
numerical experiment results that there is very limited reconnection occurring between two flux systems that are nearly
parallel to each other, and none of the associated reconnection signatures are detected, such as local brightenings and high-speed outflows of the hot plasma.

\acknowledgments{
We wish to thank the unknown referee for various comments that helped us to improve the manuscript significantly. The data are used courtesy of the NVST and \emph{SDO} science teams. \emph{SDO} is a mission of
NASA's Living With a Star Program.
This work is supported by the National Natural Science Foundations of China (11533008, 11790304, 11773039, 11673035, 11673034, 11873059, and 11790300), and Key Programs of the Chinese Academy of Sciences (QYZDJ-SSW-SLH050).
}

{}
\clearpage

\begin{figure}
\centering
\includegraphics [width=0.96\textwidth]{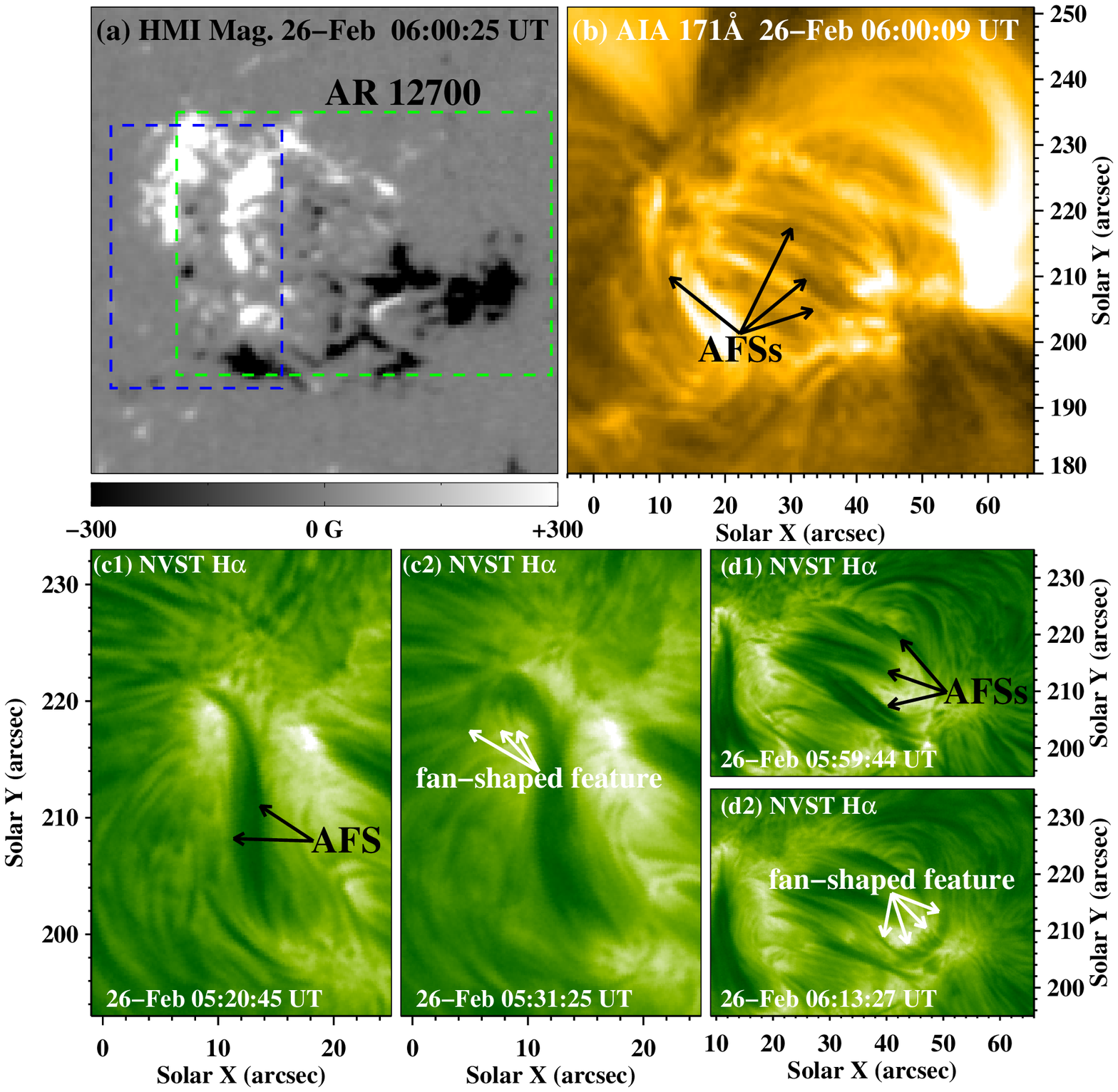}
\caption{\emph{SDO}/HMI line-of-sight (LOS) magnetogram (panel (a)), \emph{SDO}/AIA 171 {\AA} image (panel (b)) and NVST H$\alpha$ images (panels (c1)-(d2)) displaying the AR 12700 and fan-shaped activities on 2018 February 26. The blue and green rectangles in the panel (a1) indicate the FOV of the FSA1 (panels (c1)-(c2)) and that of FSA2 (panels (d1)-(d2)), respectively. The black and white arrows indicate several AFSs and the fan-shaped features, respectively.
}
\label{fig1}
\end{figure}

\begin{figure}
\centering
\includegraphics [width=0.96\textwidth]{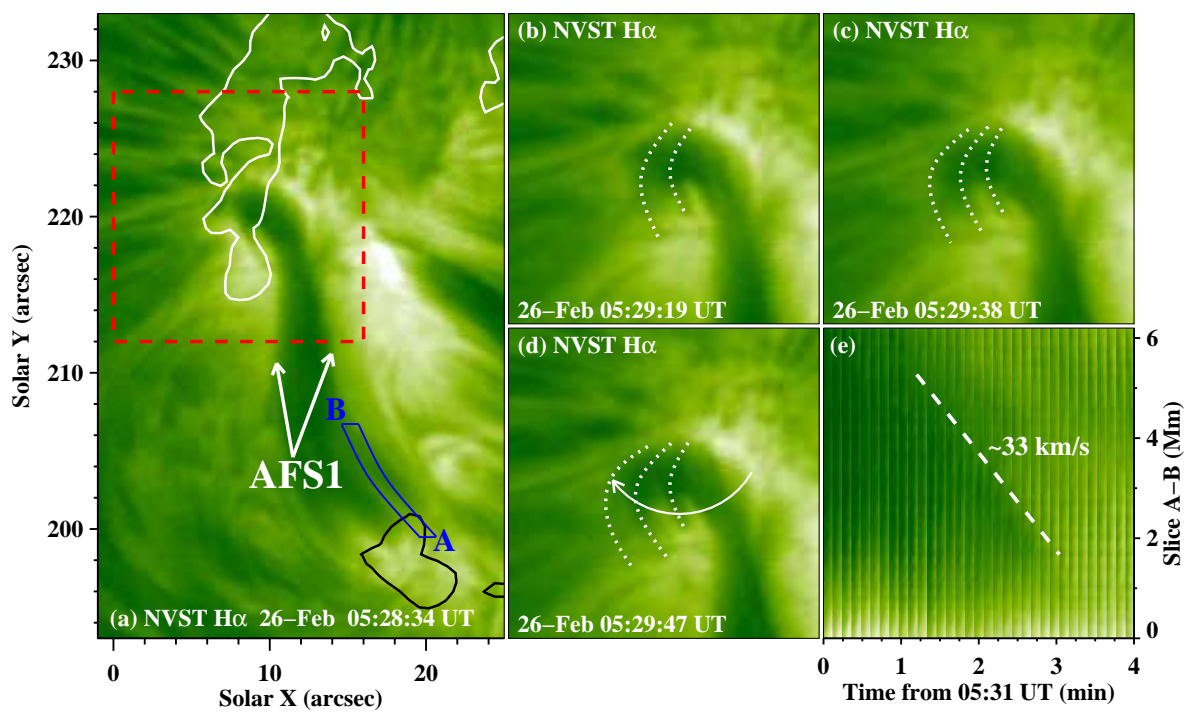}
\caption{NVST H$\alpha$ images (panels (a)-(d)) and time-slice plot (panel (e)) displaying the first fan-shaped activity at around 05:30 UT on February 26. 
The black and white curves in panel (a) are contours of the negative and positive polarities at $-$200 G and +200 G levels, respectively. The red rectangle in the panel (a) indicates the FOV of panels (b)-(d).
The blue arc-sector domain ``A-B" delineates a dark thread of the AFS1 which was used to make the time-slice plot in panel (e).
The white dotted curves in panels (b)-(d) delineate the threads that make up the first fan-shaped feature.
The white arrow in panel (d) denotes the direction of the sweeping motion of the threads. The white dashed line in panel (e) indicates the velocity of the downflow.
Online animation (movie1.mov) displays the first fan-shaped activity at H$\alpha$ wavelength shown in Figure 2. The 8 s animation covers 25 minutes from 05:15 UT to 05:40 UT on 2018 February 26.
}
\label{fig2}
\end{figure}

\begin{figure}
\centering
\includegraphics [width=0.96\textwidth]{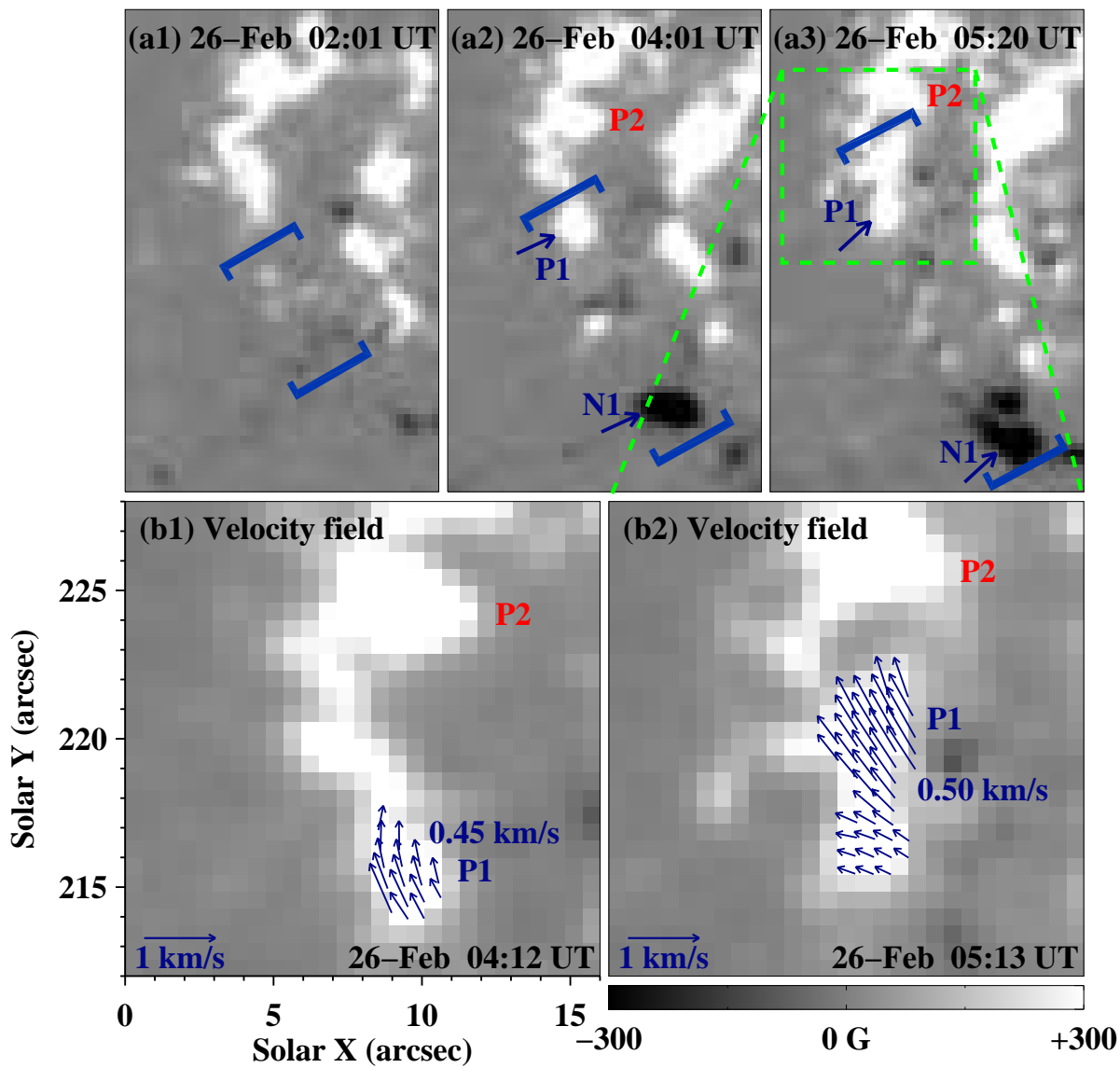}
\caption{Sequence of \emph{SDO}/HMI magnetograms displaying the evolution of the magnetic field focusing on the region of AFS1 from 02:00 UT to 05:30 UT on 2018 February 26.
In panels (a1)-(a3), the blue square brackets highlight the newly-emerging bipole connected by AFS1. The blue arrows denote the positive and negative magnetic elements of the bipole, which are also labeled by ``P1/N1". ``P2" denotes the north magnetic element with the positive polarity. The green rectangle in panel (a3) indicates the FOV of panels (b1)-(b2). In panels (b1)-(b2), the blue arrows represent the horizontal photospheric velocity fields, which are only superposed on the regions where the field strength is stronger than 300 G.
Online animation (movie2.mov) displays the photospheric magnetic fields evolution focusing on AFS1 shown in Figure 3. The 16 s animation covers 4 hr from 02:00 UT to 06:00 UT on 2018 February 26.
}
\label{fig3}
\end{figure}

\begin{figure}
\centering
\includegraphics [width=0.96\textwidth]{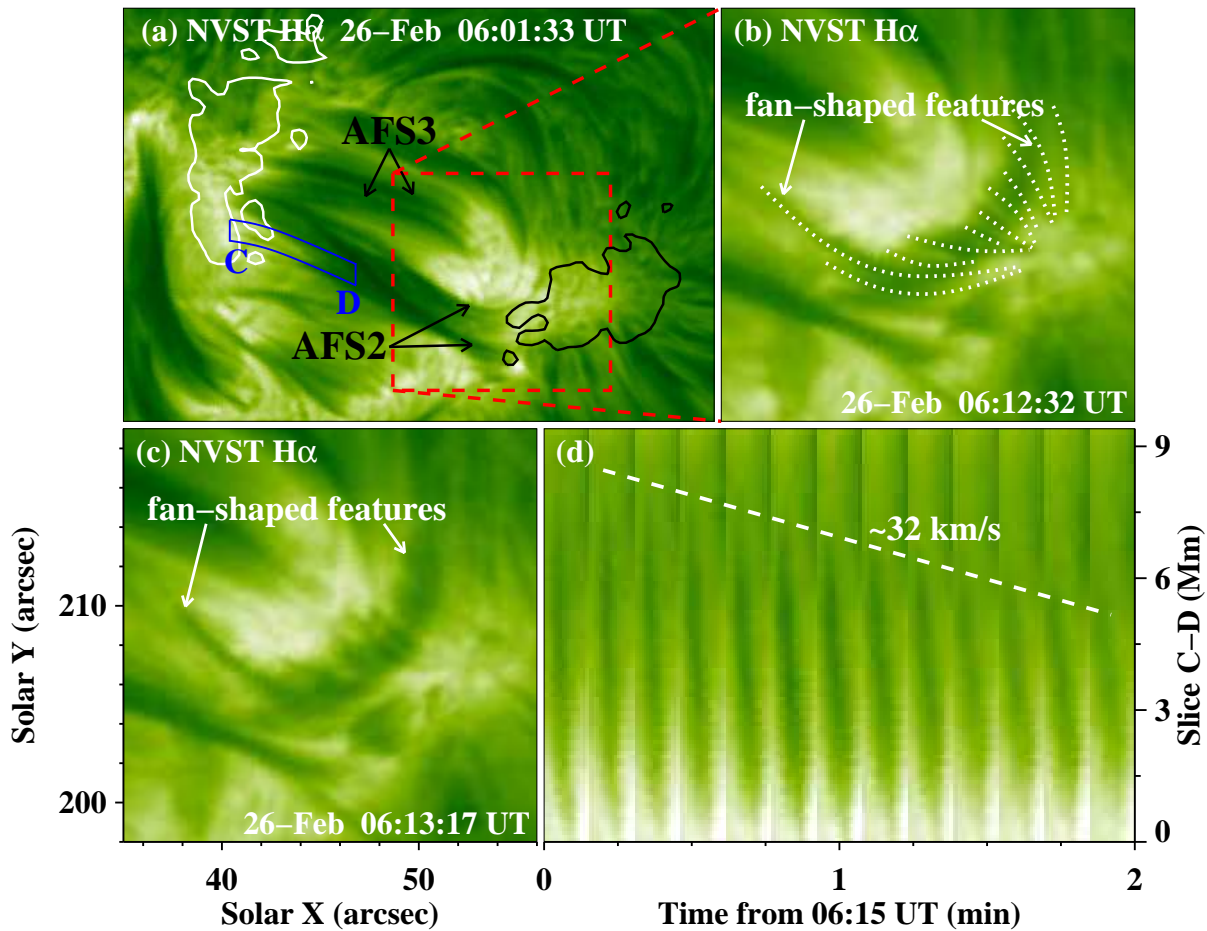}
\caption{NVST H$\alpha$ images from 06:00 UT to 06:15 UT on February 26 displaying the second fan-shaped activity. The black and white curves are contours of the negative and positive polarities at $-$150 G and +150 G levels, respectively.
The black arrows label the second AFS (AFS2) and the third AFS (AFS3). The red rectangle in panel (a) indicates the FOV of panels (b)-(c).
The white dotted curves in panels (b) delineate the threads that make up the second fan-shaped feature.
The blue arc-sector domain ``C-D"
covers a dark thread of AFS2 and was used to made the time-slice plot in panel (d). The white dashed line in panel (d) indicates the velocity of the downflow.
Online animation (movie3.mov) displays the second fan-shaped activity at H$\alpha$ wavelength shown in Figure 4. The 12 s animation covers 38 minutes from 05:58 UT to 06:36 UT on 2018 February 26.
}
\label{fig4}
\end{figure}

\begin{figure}
\centering
\includegraphics [width=0.96\textwidth]{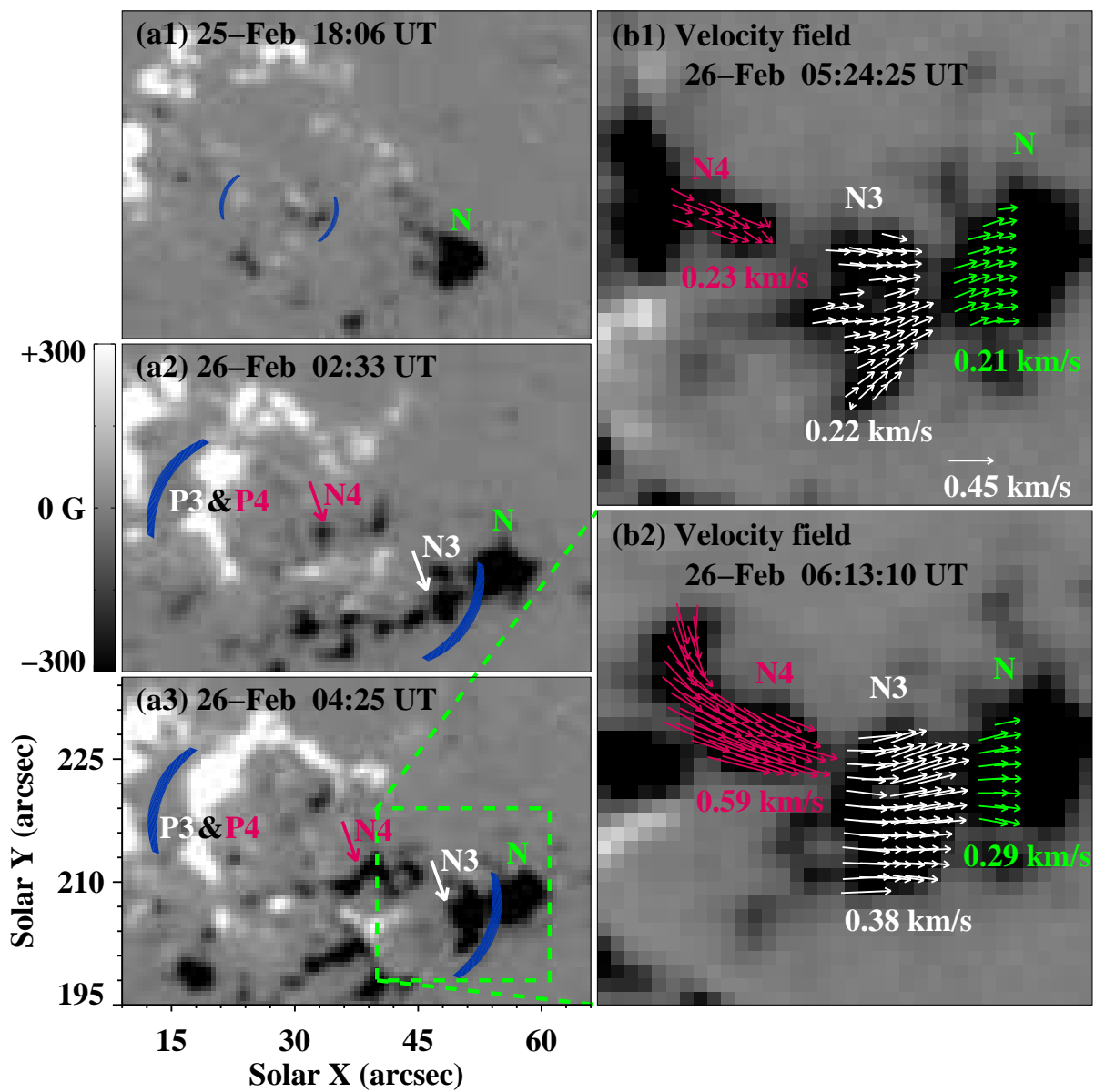}
\caption{Sequence of \emph{SDO}/HMI magnetograms displaying the magnetic evolution focusing on the AR. In panels (a1)-(a3), the blue parentheses highlight the bipole connected by the AFS2. The emerging negative magnetic patches are denoted by arrows with different colors.
The green rectangle in panel (a3) indicates the FOV of panels (b1)-(b2). In panels (b1)-(b2), the colored arrows represent the velocity fields which are only plotted on regions where the field strength is stronger than $-$ 300 G.
Online animation (movie4.mov) displays the photospheric magnetic fields evolution focusing on the target region shown in Figure 5. The 25 s animation covers 12.5 hr from 18:00 UT on February 25 to 06:32 UT on 2018 February 26.
}
\label{fig5}
\end{figure}

\begin{figure}
\centering
\includegraphics [width=0.96\textwidth]{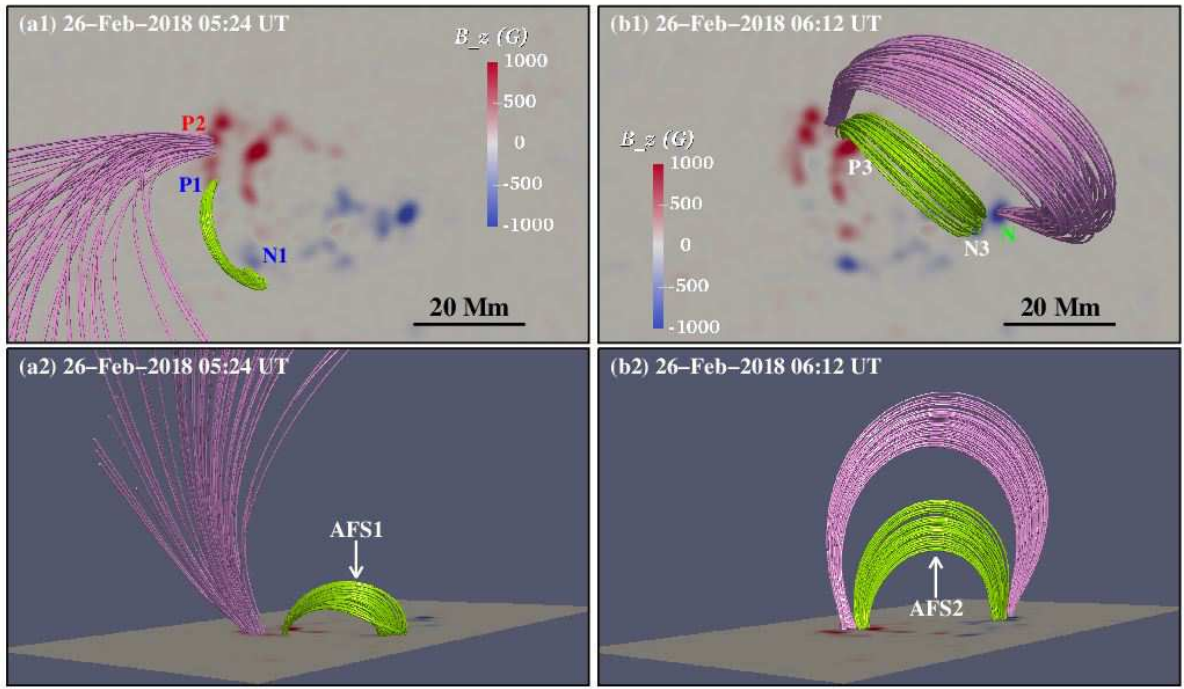}
\caption{Magnetic structures revealed by the NLFFF extrapolations for FSA1 (panels (a1)-(a2)) and FSA2 (panels (b1)-(b2)). The photospheric vertical magnetograms (Bz) are shown as the background.
The upper panels show the top view of the structures, with north as up and west to the right. The lower panels correspond to the side view.
The green curves represent the magnetic connectivity of AFSs (AFS1 and AFS2) and the pink curves represent the magnetic fields of P2 and N.
}
\label{fig6}
\end{figure}

\begin{figure}
\centering
\includegraphics [width=0.96\textwidth]{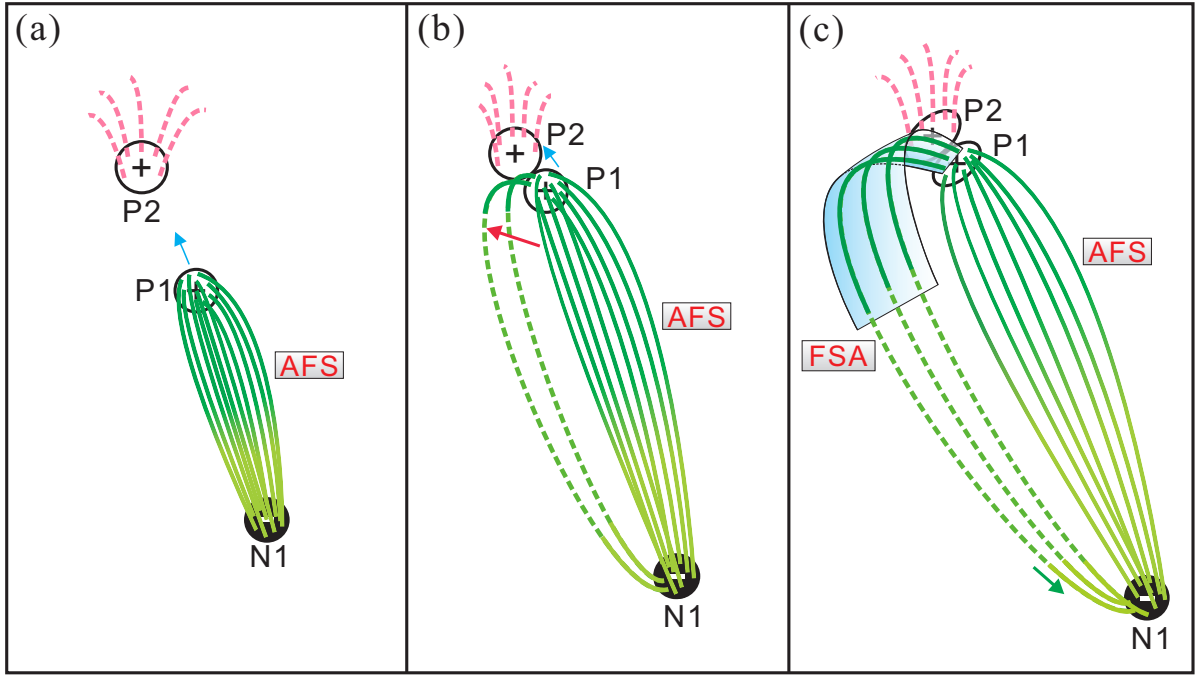}
\caption{Cartoon model illustrating the driver mechanism of the first fan-shaped activity shown in Figure 2.
Panel (a): The magnetic field configuration before the collision. ``P1/N1" denotes the positive or negative polarity connected by AFS1 and ``P2" denotes the north patch with the positive polarity as shown in Figure 3(b1).
The green solid curves represent the magnetic connectivity of AFS1 and the pink dashed curves represent the magnetic fields of P2.
The blue arrow indicates the shifting direction of P1.
Panel (b): The initiation of fan-shaped activity due to the collision between the magnetic elements with the same polarity. The red arrow denotes the lifting direction of the field lines. Panel (c): The peak of the FSA1 as shown in Figure 2(b). The solid curves on the light-blue surface represent the threads that make up the fan-shaped feature as described in Section 3. The green arrow denotes the direction of the plasma downward motion.
}
\label{fig7}
\end{figure}
\end{document}